\documentstyle[12pt]{article}
\newlength{\extraspace}
\newlength{\extraspaces}
\newcommand{\be}{\begin{equation}
\addtolength{\abovedisplayskip}{\extraspaces}
\addtolength{\belowdisplayskip}{\extraspaces}
\addtolength{\abovedisplayshortskip}{\extraspace}
\addtolength{\belowdisplayshortskip}{\extraspace}}
\newcommand{\ee}{\end{equation}}

\newcommand{\ba}{\begin{eqnarray}
\addtolength{\abovedisplayskip}{\extraspaces}
\addtolength{\belowdisplayskip}{\extraspaces}
\addtolength{\abovedisplayshortskip}{\extraspace}
\addtolength{\belowdisplayshortskip}{\extraspace}}
\newcommand{\ea}{\end{eqnarray}}

\newcommand{\newsection}[1]{
\vspace{15mm}
\pagebreak[3]
\addtocounter{section}{1}
\setcounter{equation}{0}
\setcounter{subsection}{0}
\setcounter{footnote}{0}
\begin{flushleft}
{\large\bf \thesection. #1}
\end{flushleft}
\nopagebreak
\medskip
\nopagebreak}

\newcommand{\Tr}{{\rm Tr}}

\newcommand{\Dmrns}{{\cal D}_{\mu\rho,\nu\sigma}}

\begin{document}

\addtolength{\baselineskip}{.8mm}

{\thispagestyle{empty}
\noindent \hspace{1cm}  \hfill HD--THEP--98--34 \hspace{1cm}\\
\mbox{}                 \hfill November 1998 \hspace{1cm}\\

\begin{center}\vspace*{1.0cm}
{\large\bf Field strength correlators in QCD:} \\
{\large\bf new fits to the lattice data} \\
\vspace*{1.0cm}
{\large Enrico Meggiolaro\footnote{E--mail: 
E.Meggiolaro@thphys.uni-heidelberg.de} } \\
\vspace*{0.5cm}{\normalsize
{Institut f\"ur Theoretische Physik, \\
Universit\"at Heidelberg, \\
Philosophenweg 16, \\ 
D--69120 Heidelberg, Germany.}} \\
\vspace*{2cm}{\large \bf Abstract}
\end{center}

\noindent
We discuss the results obtained by fitting the lattice data of the
gauge--invariant field strength correlators in QCD with some particular
functions which are commonly used in the literature in some
phenomenological approaches to high--energy hadron--hadron scattering.
A comparison is done with the results obtained in the original fits to
the lattice data.\\
\vspace{1.0cm}
\noindent
(PACS code: 12.38.Gc)
}
\vfill\eject

\newsection{Introduction}

\noindent
The gauge--invariant two--point correlators of the field strengths in the 
QCD vacuum are defined as
\be
\Dmrns(x) = g^2 \langle 0| 
\Tr \left\{ G_{\mu\rho}(0) S(0,x) G_{\nu\sigma}(x) S^\dagger(0,x) \right\}
|0\rangle ~,
\ee
where $G_{\mu\rho} = T^a G^a_{\mu\rho}$ and $T^a$ are the matrices of the 
algebra of the colour group SU($N_c$) in the fundamental representation 
(For $N_c = 3$, $T^a = \lambda_a/2$, where $\lambda_a$ are the Gell--Mann 
matrices). The trace in (1.1) is taken with respect of the colous indices. 
Moreover, in Eq. (1.1),
\be
S(0,x) \equiv {\rm P} \exp \left( i g \int_0^x dz^\mu A_\mu (z) \right) ~,
\ee
with $A_\mu = T^a A^a_\mu$, is the Schwinger phase operator needed to 
parallel--transport the tensor $G_{\nu\sigma}(x)$ to the point $0$.
``P'' stands for ``path ordering'': for simplicity, we take $S(0,x)$ along 
the straight--line path from $0$ to $x$. 

These field--strength correlators play an important role in hadron physics.
In the spectrum of heavy $Q \bar{Q}$ bound states, they govern the effect 
of the gluon condensate on the level splittings 
\cite{Gromes82,Campostrini86,Simonov95}.
They are the basic quantities in models of stochastic confinement of colour
\cite{Dosch87,Dosch88,Simonov89}
and in the description of high--energy hadron scattering
\cite{Nachtmann84,Landshoff87,Kramer90,Dosch94}.
In some recent works \cite{Dorokhov97,Ilgenfritz97}, these correlators have 
been semi--classically evaluated in the single--instanton approximation and
in the instanton dilute--gas model, so providing useful information about the 
role of the semiclassical modes forming the QCD vacuum.

In the Euclidean theory, translational, $O(4)$-- and parity invariance
require the correlator (1.1) to be of the following form 
\cite{Dosch87,Dosch88,Simonov89}:
\ba
\lefteqn{
\Dmrns(x) = (\delta_{\mu\nu}\delta_{\rho\sigma} - 
\delta_{\mu\sigma}\delta_{\rho\nu})
\left[ {\cal D}(x^2) + {\cal D}_1(x^2) \right] } \nonumber \\
& & + (x_\mu x_\nu \delta_{\rho\sigma} - x_\mu x_\sigma \delta_{\rho\nu} 
+ x_\rho x_\sigma \delta_{\mu\nu} - x_\rho x_\nu \delta_{\mu\sigma})
{\partial{\cal D}_1(x^2) \over \partial x^2} ~,
\ea
where ${\cal D}$ and ${\cal D}_1$ are invariant functions of $x^2$.

These functions ${\cal D}(x^2)$ and ${\cal D}_1(x^2)$ have been directly 
determined (in the Euclidean theory) by numerical simulations on a lattice in 
the {\it quenched} (i.e., pure gauge) theory, with gauge--group SU(2) 
\cite{Campostrini84}, in the {\it quenched} SU(3) theory in the range of 
physical distances between 0.1 and 1 fm \cite{DiGiacomo92,npb97} and also in 
full QCD, i.e., including the effects of dynamical fermions \cite{plb97}.
In another approach \cite{Brambilla98}, they have been extracted (in the
{\it quenched} SU(3) theory) from lattice calculations of the heavy--quark
potential, under the assumptions of the model of the stochastic vacuum
\cite{Dosch87,Dosch88,Simonov89}.

It is convenient to define a ${\cal D}_\parallel(x^2)$ and a 
${\cal D}_\perp(x^2)$ as follows:
\ba
{\cal D}_\parallel &\equiv& {\cal D} + {\cal D}_1 + x^2 {\partial{\cal D}_1
\over \partial x^2} ~, \nonumber \\
{\cal D}_\perp &\equiv& {\cal D} + {\cal D}_1 ~.
\ea
In Figs. 1--2 we display the results for ${\cal D}_\parallel/\Lambda^4_L$ 
and ${\cal D}_\perp/\Lambda^4_L$ versus the physical distance in fermi units,
obtained in the {\it quenched} (i.e., pure--gauge) theory, with gauge
group SU(3): data are taken from Refs. \cite{DiGiacomo92,npb97}.
$\Lambda_L$ is the fundamental constant of QCD in the lattice 
renormalization scheme, in the pure--gauge case: its value, extracted from 
the string tension \cite{Michael88}, turns out to be about 4.92 MeV for the 
gauge group SU(3) and in the absence of quarks.

In Fig. 3 we display the results for ${\cal D}_\parallel/\Lambda^4_F$ 
and ${\cal D}_\perp/\Lambda^4_F$ versus the physical distance in fermi units,
obtained in full QCD, with gauge group SU(3) and $N_f = 4$ flavours of
{\it staggered} fermions with a quark mass $m_q = 0.01$ in lattice units:
data are taken from Ref. \cite{plb97}.
In Fig. 4 we display the corresponding results for the quark mass 
$m_q = 0.02$ \cite{plb97}.
$\Lambda_F$ is an effective $\Lambda$--parameter for QCD in the 
lattice renormalization scheme, for the gauge group SU(3) and $N_f = 4$ 
flavours of quarks. It was defined in Ref. \cite{plb97}, where the value
$\Lambda_F \simeq 1.07$ MeV was derived and used.

In Refs. \cite{DiGiacomo92,npb97,plb97} best fits to the lattice data
with the functions
\ba
{\cal D}(x^2) &=& A_0 {\rm e}^{-|x|/\lambda_A}
+ {a_0 \over |x|^4} {\rm e}^{-|x|/\lambda_a} ~,
\nonumber \\
{\cal D}_1(x^2) &=& A_1 {\rm e}^{-|x|/\lambda_A}
+ {a_1 \over |x|^4} {\rm e}^{-|x|/\lambda_a} ~,
\ea
were performed: the results obtained for all the various cases are 
reported in Table I. From these results one sees that, in order to obtain
a fit with a good $\chi^2/N$, it is necessary
to include a perturbative--like term behaving as $1/|x|^4$ (indeed,
a term of this form is predicted by ordinary perturbation theory; see
for example Ref. \cite{Jamin98} and references therein)
in addition to the nonperturbative exponential term in the parametrization
for both ${\cal D}$ and ${\cal D}_1$. In Table I we also report the results
from the fits to the restricted set of data of the correlators obtained
in Ref. \cite{DiGiacomo92}, corresponding to physical distances
$|x| \ge 0.4$ fm: in this case a fit with only the nonperturbative
exponential terms for ${\cal D}$ and ${\cal D}_1$ turns out to be
acceptable.

In this paper we report the results obtained by fitting the lattice 
data of the gauge--invariant field strength correlators in QCD, displayed in
Figs. 1--4, with some particular functions which are commonly used in the 
literature in some phenomenological approaches to high--energy hadron--hadron 
scattering. Therefore, we think that these results may be useful to those
people working with this alternative parametrization.
In Sect. 2 we give a brief review on the new parametrization used 
for the best fits and the results so obtained are discussed.
In Sect. 3 some quantities of physical interest are extracted from our
best fits and compared with the corresponding results obtained in 
the original fits to the lattice data.

\newsection{New fits to the lattice data}

\noindent
In this section we discuss the results obtained by fitting the lattice data
reported in Figs. 1--4 with the following functions:
\ba
{\cal D}(x^2) &=& {\pi^2 \over 6} \kappa G_2 \tilde{\cal D}(x^2) 
+ {a_0 \over |x|^4} {\rm e}^{-|x|/\lambda_a} ~, \nonumber \\
{\cal D}_1(x^2) &=& {\pi^2 \over 6} (1-\kappa) G_2 \tilde{\cal D}_1(x^2) 
+ {a_1 \over |x|^4} {\rm e}^{-|x|/\lambda_a} ~,
\ea
where $\tilde{\cal D}(x^2)$ and $\tilde{\cal D}_1(x^2)$ are
so defined:
\ba
\tilde{\cal D}(x^2) &=& \left( {3\pi |x| \over 8a} \right) 
K_1 \left( {3\pi |x| \over 8a} \right) - {1 \over 4}
\left( {3\pi |x| \over 8a} \right)^2
K_0 \left( {3\pi |x| \over 8a} \right) ~, \nonumber \\
\tilde{\cal D}_1(x^2) &=& \left( {3\pi |x| \over 8a} \right) 
K_1 \left( {3\pi |x| \over 8a} \right) ~.
\ea
$K_0$ and $K_1$ are the modified Bessel functions. These expressions
for the ``nonperturbative'' parts of ${\cal D}$ and ${\cal D}_1$, i.e,
the expressions reported in Eqs. (2.1) and (2.2), with the exclusion
of the ``perturbative--like'' $1/|x|^4$ pieces, are extensively used in the 
literature in many phenomenological approaches to high--energy hadron--hadron 
scattering (see for example Refs. \cite{Dosch94,Nachtmann97,Nachtmann98}
and references therein). They were proposed for the first time in Ref. 
\cite{Dosch94}, where also a preliminary fit to lattice data was performed.
However, at that time only lattice data obtained in the {\it quenched}
SU(3) theory, in the range of physical distances between 0.4 and 1 fm,
were available \cite{DiGiacomo92}.
Therefore, we think that a re--analysis of this parametrization, fitting
also the new lattice data now available in the {\it quenched} SU(3) theory 
in the range of physical distances between 0.1 and 1 fm \cite{npb97} and in 
full QCD \cite{plb97}, may be useful for the practitioners in this field.

Before discussing the results of the fits, we shall remind the reader
of some technical details about the parametrization (2.1).
The functions $\tilde{\cal D}(x^2)$ and $\tilde{\cal D}_1(x^2)$ are
normalized to 1 in $x=0$:
\be
\tilde{\cal D}(0) = \tilde{\cal D}_1(0) = 1 ~.
\ee
As we shall discuss in the next section, this implies that the parameter
$G_2$ in Eq. (2.1) should be identified with the gluon condensate.
The parameter $\kappa$ measures the non--Abelian character of the
correlator: in fact, $\kappa = 0$ in an Abelian theory, if there are
no magnetic monopoles, while there is no reason for the ${\cal D}$--term
to vanish in a non--Abelian theory.
The expression for $\tilde{\cal D}(x^2)$ in Eq. (2.2) comes from
the following {\it ansatz}:
\be
\tilde{\cal D}(x^2) = {27 \over 64} a^{-2} \displaystyle\int {\rm d}^4 k
{\rm e}^{ikx} {k^2 \over \left[ k^2 + \left( {3\pi \over 8a} \right)^2
\right]^4} ~,
\ee
where the length--scale $a$ enters into this parametrization as a 
``correlation length'', being defined as
\be
a \equiv \displaystyle\int_0^{+\infty} {\rm d}|x| \tilde{\cal D}(x^2) ~.
\ee
The correlation function $\tilde{\cal D}(x^2)$ is negative at large
distances, with the following asymptotic behaviour:
\be
\tilde{\cal D}(x^2) \mathop\sim_{x^2 \to \infty} 
-{1 \over 4} \sqrt{\pi \over 2} \left( {3\pi |x| \over 8a} \right)^{3 \over 2}
\exp \left( -{3\pi |x| \over 8a} \right) ~.
\ee
The function $\tilde{\cal D}_1(x^2)$ is chosen such that
\be
\left( 4 + x_\mu {\partial \over \partial x_\mu} \right)
\tilde{\cal D}_1(x^2) = 4 \tilde{\cal D}(x^2) ~,
\ee
which leads to
\be
\tilde{\cal D}_1(x^2) = {1 \over |x|^4} \displaystyle\int_0^{x^2}
{\rm d}z^2 \tilde{\cal D}(z^2) ~,
\ee
and then to the expression in Eq. (2.2).

The results of the fits to the lattice data using the parametrization
(2.1)--(2.2) for ${\cal D}$ and ${\cal D}_1$ are reported in Table II.
The continuum lines in Figs. 1--4 have been obtained using the parameters of 
these best fits, in the cases where all the parameters were free.
The dashed lines in Figs. 1--2 correspond to the nonperturbative parts
only in Eq. (2.1) and are derived using the same parameters used for the
corresponding continuum lines. They are drawn in this particular case
({\it quenched} SU(3) theory) in order to illustrate the role of the
perturbative--like terms.

From the results in Table II, one sees that, in order to obtain
fits with a good $\chi^2/N$, it is necessary to include the perturbative--like 
terms $\sim 1/|x|^4$ in the parametrization for ${\cal D}$ and ${\cal D}_1$.
In fact one also sees directly from Figs. 1--2 that these terms are necessary
to well describe the behaviour of the correlators at small distances (down
to 0.1 fm), while they are less important in the range of distances between
0.4 and 1 fm.
The coefficients $a_0$ and $a_1$ turns out to be comparable with (even if
slightly smaller than) the coefficients derived in the corresponding cases
in Table I, obtained using the parametrization (1.5).
Even restricting the set of {\it quenched} data to those obtained in 
Ref. \cite{DiGiacomo92}, corresponding to physical distances $|x| \ge 0.4$ fm, 
one finds that a fit with only the nonperturbative terms in Eq. (2.1), i.e., 
fixing $a_0 = a_1 = 0$, is more acceptable (when compared to the same fit 
applied to the entire set of data between 0.1 and 1 fm), but the $\chi^2/N$ 
is still too high ($\sim 3.8$).
When all the parameters in the fit are free, the $\chi^2/N$ turns out to
be acceptable in all the various cases examined, even if it is
systematically a bit larger than the $\chi^2/N$ obtained in the corresponding
cases examined in Table I, using the original parametrization (1.5).

Therefore, we can conclude that the expressions (2.1)--(2.2) are a good
parametrization of the correlators ${\cal D}$ and ${\cal D}_1$, in the
range of physical distances where the lattice data are available (i.e.,
0.1--1 fm for {\it quenched} QCD and 0.3--1 fm for full QCD).
However, the parametrization (1.5) appears to be slightly preferable.
In the next section we shall discuss some quantities of physical interest 
which can be extracted from the results of the best fits in Tables I and II.

\newsection{Discussion}

\noindent
Three quantities of physical interest can be extracted from our fits to the
lattice data:
\begin{itemize}
\item[(1)] The correlation length $l_G$ of the gluon field strengths, 
defined as (``$np$'' stands for ``nonperturbative'', in the sense explained 
in the previous sections)
\be
l_G \equiv {1 \over {\cal D}^{(np)}(0)} \int_0^{+\infty} 
{\rm d}|x| {\cal D}^{(np)}(x^2) ~. 
\ee
\item[(2)] The so--called ``gluon condensate'', defined as
\be
G_2 \equiv \langle {\alpha_s \over \pi} :G^a_{\mu\nu} G^a_{\mu\nu}: \rangle
~~~~~~~~~ (\alpha_s = {g^2 \over 4\pi}) ~.
\ee
\item[(3)] The parameter $\kappa$, defined as 
\be
\kappa = { {\cal D}^{(np)}(0) \over {\cal D}^{(np)}(0) 
+ {\cal D}^{(np)}_1(0) } ~.
\ee
\end{itemize}
The results obtained are summarized in Table III. 

The quantities  $l_G$ and $G_2$ play an important role in phenomenology. 
The correlation length is relevant for the description of vacuum models
\cite{Dosch87,Dosch88,Simonov89}.
The relevance of the gluon condensate was first pointed out by Shifman, 
Vainshtein and Zakharov (SVZ) \cite{SVZ79}. 
It is a fundamental quantity in QCD, in the context of the SVZ sum rules.

The physical meaning of the $\kappa$ parameter has been already discussed
in the previous section: it measures the non--Abelian character of the
correlator, since one expects that $\kappa = 0$ in an Abelian theory with
no magnetic monopoles present.
This parameter appears explicitely in the parametrization (2.1),
by virtue of Eq. (2.3). Instead, using the parametrization (1.5) one
finds the following expression for $\kappa$:
\be
\kappa = {A_0 \over A_0 + A_1} ~.
\ee
From the results reported in Table III, one sees that, in both 
parametrizations (1.5) and (2.1)--(2.2), the parameter
$\kappa$ appears to decrease when increasing the quark mass, tending
towards the pure--gauge value (obviously, when evaluating the field
strength correlators (1.1), the {\it quenched}, i.e., pure--gauge,
limit coincides with the large quark--mass limit, $m_q \to \infty$).
In other words, the non--Abelian character of the correlator ${\cal D}$
appears to increse when approaching the chiral limit $m_q \to 0$.
The values of $\kappa$ obtained using the parametrization (2.1)--(2.2)
in all the cases examined are a bit larger than the corresponding values of 
$\kappa$ obtained using the parametrization (1.5).

Now let us discuss the results for the correlation length $l_G$.
Using the parametrization (1.5) one easily finds $l_G 
= \lambda_A$, while in the parametrization (2.1)--(2.2) one has $l_G = a$,
according to Eqs. (2.5) and (2.3).
From the results in Table III, one sees that both $\lambda_A$ and $a$ 
decrease by increasing the quark mass, when going from chiral to 
{\it quenched} QCD. 
Obviously, the difference between $\lambda_A$ and $a$ is due to the 
different parametrization used for the correlators.

Now we come to the gluon condensate. As pointed out in Ref. \cite{plb97}, 
the lattice provides us with a regularized determination of the correlators.
We shall briefly repeat here the argumentation originally reported in
Ref. \cite{plb97}, for the benefit of the reader.
At small distances $x$ a Wilson operator--product--expansion (OPE) 
\cite{Wilson69} is expected to hold.
The regularized correlators will then mix to the identity operator {\bf 1},
to the renormalized local operators of dimension four,
${\alpha_s \over \pi}$:$G^a_{\mu\nu} G^a_{\mu\nu}$: and 
$m_f$:$\bar{q}_f q_f$: ($f = 1, \ldots, N_f$, $N_f$ being the number of 
quark flavours), and to operators of higher dimension:
\be
{1 \over 2\pi^2} {\cal D}_{\mu\nu,\mu\nu}(x) \mathop\sim_{x\to0}
C_{\bf 1}(x) \langle {\bf 1} \rangle + C_g(x) G_2 + 
\displaystyle\sum_{f=1}^{N_f} C_f(x) m_f \langle :\bar{q}_f q_f: \rangle
+ \ldots ~.
\ee
The mixing to the identity operator $C_{\bf 1}(x)$ shows up as a $c/|x|^4$ 
behaviour at small $x$. The mixings to the operators of dimension four 
$C_g(x)$ and $C_f(x)$ are expected to behave as constants for $x \to 0$, 
while the other Wilson coefficients in the OPE (3.5) are expected to vanish 
when $x \to 0$ (for dimensional reasons).
The coefficients of the Wilson expansion are usually determined in 
perturbation theory and are known to be plagued by the so--called 
``infrared renormalons'' (see for example Ref. \cite{Mueller85} and 
references therein). 
In the same spirit of Ref. \cite{plb97}, we shall assume that the renormalon 
ambiguity can be safely neglected in the extrapolation for $x \to 0$ of our 
correlators.
With the normalization of Eq. (3.5), this gives $C_g(0) \simeq 1$.
On the same line, the contribution from the quark operators in (3.5) can be 
neglected, for the reasons explained in Ref. \cite{plb97}.
Within these approximations, one immediately recognizes that the parameter
$G_2$ in the parametrization (2.1)--(2.2), with the normalization condition
(2.3), coincides with the gluon condensate as defined in (3.2).
Moreover, when using the parametrization (1.5) for the correlators,
one obtains the following expression for the gluon condensate:
\be
G_2 \simeq {6 \over \pi^2} (A_0 + A_1) ~.
\ee
For both parametrizations (1.5) and (2.1)--(2.2), the gluon condensate $G_2$ 
appears to increase with the quark mass, as expected \cite{NSVZ81}, tending 
towards the asymptotic (pure--gauge) value. However, the values of $G_2$
extracted from the fits using the parametrization (2.1)--(2.2) are more than 
a factor two smaller than the corresponding values of $G_2$ extracted from 
the fits using the parametrization (1.5), in all the cases examined.

\bigskip
\noindent {\bf Acknowledgements}
\smallskip
 
I would like to thank Prof. Otto Nachtmann for useful discussions
and in particular for his encouragement to write this paper. 

\vfill\eject

{\renewcommand{\Large}{\normalsize}
}

\vfill\eject

\noindent
\begin{center}
{\bf TABLE CAPTIONS}
\end{center}
\vskip 0.5 cm
\begin{itemize}
\item [\bf Tab.~I.] Results obtained from a best fit to the data of the 
gauge--invariant field strength correlators with the functions (1.5), 
in the various cases that we have examined: 
``q'' stands for {\it ``quenched''} data in the range of physical
distance between 0.1 and 1 fm \cite{npb97};
``q*'' stands for {\it ``quenched''} data in the range of physical
distance between 0.4 and 1 fm \cite{DiGiacomo92};
``f (I)'' and ``f (II)'' stand for ``full--QCD'' data with quark mass 
(in lattice units) 0.01 and 0.02 respectively \cite{plb97}.
An asterisk $(*)$ near the value of some parameter means that the parameter
was fixed to that value. 
\bigskip
\item [\bf Tab.~II.] Results obtained from a best fit to the data of the
gauge--invariant field strength correlators with the functions (2.1)--(2.2),
in the various cases that we have examined: the notation used is the
same as in Table I.
\bigskip
\item [\bf Tab.~III.] The values of some quantities of physical interest
extracted from the best--fit results in Tables I (``exp'') and II (``bessel'') 
in the cases where all the parameters were free. The notation used is the
same as in Tables I and II.
Reported errors refer only to our determination 
and do not include the uncertainty on the physical scale.
\end{itemize}

\vfill\eject

\vskip 1cm

\centerline{\bf Table I}

\vskip 5mm

\moveleft .5 in
\vbox{\offinterlineskip
\halign{\strut
\vrule \hfil\quad $#$ \hfil \quad & 
\vrule \hfil\quad $#$ \hfil \quad & 
\vrule \hfil\quad $#$ \hfil \quad & 
\vrule \hfil\quad $#$ \hfil \quad & 
\vrule \hfil\quad $#$ \hfil \quad & 
\vrule \hfil\quad $#$ \hfil \quad & 
\vrule \hfil\quad $#$ \hfil \quad & 
\vrule \hfil\quad $#$ \hfil \quad \vrule \cr
\noalign{\hrule}
{\rm theory} &
A_0/\Lambda^4 &
A_1/\Lambda^4 &
1/(\lambda_A\Lambda) &
a_0 &
a_1 &
1/(\lambda_a\Lambda) &
\chi^2/N \cr
& \times 10^{-8} & \times 10^{-8} & & & & & \cr
\noalign{\hrule}
\noalign{\hrule}
{\rm q} & 3.34(20) & 0.70(10) & 182(3) &  
0.69(6) &  0.46(3) &  94(15) & 1.7 \cr
\noalign{\hrule}
{\rm q} & 8.37(22) & 2.94(9) & 233(2) &
0~(*) & 0~(*) & 0~(*) & 33 \cr
\noalign{\hrule}
{\rm q*} & 3.11(61) & 0.83(23) & 183(8) & 
0.22(12) & 0.12(7) & 0(343) & 1.4 \cr
\noalign{\hrule}
{\rm q*} & 3.62(19) & 1.23(7) & 183(3) & 
0~(*) & 0~(*) & 0~(*) & 1.3 \cr
\noalign{\hrule}
{\rm f~(I)} & 174(24) & 20(10) & 544(27) &
0.71(3) & 0.45(3) & 42(11) & 0.5 \cr
\noalign{\hrule}
{\rm f~(I)} & 438(17) & 303(9) & 642(8) &
0~(*) & 0~(*) & 0~(*) & 51 \cr
\noalign{\hrule}
{\rm f~(II)} & 348(42) & 46(21) & 631(23) & 
0.66(3) & 0.39(3) & 61(20) &  0.7 \cr
\noalign{\hrule}
{\rm f~(II)} & 734(21) & 354(10) & 713(7) & 
0~(*) & 0~(*) & 0~(*) & 27 \cr
\noalign{\hrule}
}}

\vskip 1cm

\centerline{\bf Table II}

\vskip 5mm

\moveleft .5 in
\vbox{\offinterlineskip
\halign{\strut
\vrule \hfil\quad $#$ \hfil \quad & 
\vrule \hfil\quad $#$ \hfil \quad & 
\vrule \hfil\quad $#$ \hfil \quad & 
\vrule \hfil\quad $#$ \hfil \quad & 
\vrule \hfil\quad $#$ \hfil \quad & 
\vrule \hfil\quad $#$ \hfil \quad & 
\vrule \hfil\quad $#$ \hfil \quad & 
\vrule \hfil\quad $#$ \hfil \quad \vrule \cr
\noalign{\hrule}
{\rm theory} &
G_2/\Lambda^4 &
\kappa &
1/(a\Lambda) &
a_0 &
a_1 &
1/(\lambda_a\Lambda) &
\chi^2/N \cr
& \times 10^{-8} & & & & & & \cr
\noalign{\hrule}
\noalign{\hrule}
{\rm q} & 0.95(5) & 0.89(2) & 122(2) &  
0.56(4) &  0.38(3) &  48(12) & 2 \cr
\noalign{\hrule}
{\rm q} & 2.36(5) & 0.82(1) & 150(1) &
0~(*) & 0~(*) & 0~(*) & 54 \cr
\noalign{\hrule}
{\rm q*} & 0.62(8) & 0.85(4) & 116(3) & 
0.50^{+0.25}_{-0.07} & 0.22^{+0.10}_{-0.04} & 0(33) & 1.5 \cr
\noalign{\hrule}
{\rm q*} & 1.14(4) & 0.80(1) & 122(1) & 
0~(*) & 0~(*) & 0~(*) & 3.8 \cr
\noalign{\hrule}
{\rm f~(I)} & 53(7) & 0.95(5) & 387(16) &
0.66(2) & 0.41(2) & 0.0(1) & 0.5 \cr
\noalign{\hrule}
{\rm f~(I)} & 185(4) & 0.60(1) & 457(4) &
0~(*) & 0~(*) & 0~(*) & 84 \cr
\noalign{\hrule}
{\rm f~(II)} & 97(10) & 0.94(3) & 436(13) & 
0.62(3) & 0.36(2) & 0.00(5) & 1 \cr
\noalign{\hrule}
{\rm f~(II)} & 284(5) & 0.73(1) & 502(3) & 
0~(*) & 0~(*) & 0~(*) & 53 \cr
\noalign{\hrule}
}}

\vskip 1cm

\centerline{\bf Table III}

\vskip 5mm

\moveright .4 in
\vbox{\offinterlineskip
\halign{\strut
\vrule \hfil\quad $#$ \hfil \quad & 
\vrule \hfil\quad $#$ \hfil \quad & 
\vrule \hfil\quad $#$ \hfil \quad & 
\vrule \hfil\quad $#$ \hfil \quad & 
\vrule \hfil\quad $#$ \hfil \quad \vrule \cr
\noalign{\hrule}
{\rm theory} &
{\rm fit} &
G_2 &
\kappa &
\lambda_A,~a \cr
& & ({\rm GeV}^4) & & ({\rm fm}) \cr
\noalign{\hrule}
\noalign{\hrule}
{\rm f~(I)} & {\rm exp} & 0.015(3) & 0.90(6) & 0.34(2) \cr
\noalign{\hrule}
{\rm f~(I)} & {\rm bessel} & 0.007(1) & 0.95(5) & 0.48(2) \cr
\noalign{\hrule}
{\rm f~(II)} & {\rm exp} & 0.031(5) & 0.88(6) & 0.29(1) \cr
\noalign{\hrule}
{\rm f~(II)} & {\rm bessel} & 0.0127(13) & 0.94(3) & 0.42(1) \cr
\noalign{\hrule}
{\rm q} & {\rm exp} & 0.144(11) & 0.83(3) & 0.220(4) \cr
\noalign{\hrule}
{\rm q} & {\rm bessel} & 0.056(3) & 0.89(2) & 0.328(5) \cr
\noalign{\hrule}
}}

\newpage

\noindent
\begin{center}
{\bf FIGURE CAPTIONS}
\end{center}
\vskip 0.5 cm
\begin{itemize}
\item [\bf Fig.~1.] The function ${\cal D}_\parallel/\Lambda^4_L$ versus the 
physical distance in fermi units, obtained in the {\it quenched} 
(i.e., pure--gauge) theory, with gauge group SU(3) ($\Lambda_L \simeq 4.92$ 
MeV). Data are taken from Refs. \cite{DiGiacomo92,npb97}, while the curves are 
obtained from our best fit with the functions (2.1)--(2.2) 
[fit no. 1 in Table II]: the continuum line corresponds to the entire
correlator, while the dashed line corresponds to its nonperturbative part only.
\bigskip
\item [\bf Fig.~2.] The same as in Fig. 1 for the function 
${\cal D}_\perp/\Lambda^4_L$.
\bigskip
\item [\bf Fig.~3.] The functions ${\cal D}_\perp/\Lambda^4_F$ (upper curve)
and ${\cal D}_\parallel/\Lambda^4_F$ (lower curve) versus the physical distance 
in fermi units, obtained in full QCD, with gauge group SU(3) and $N_f = 4$ 
flavours of {\it staggered} fermions with a quark mass $m_q = 0.01$ in 
lattice units ($\Lambda_F \simeq 1.07$ MeV).
Data are taken from Ref. \cite{plb97}, while the curves are 
obtained from our best fit with the functions (2.1)--(2.2) 
[fit no. 5 in Table II].
\bigskip
\item [\bf Fig.~4.] The same as in Fig. 3 for a quark mass $m_q = 0.02$ in 
lattice units ($\Lambda_F \simeq 1.07$ MeV).
Data are taken from Ref. \cite{plb97}, while the curves are 
obtained from our best fit with the functions (2.1)--(2.2) 
[fit no. 7 in Table II].
\end{itemize}

\vfill\eject


\begin{thebibliography}{99}
\bibitem{Gromes82}
D. Gromes, Phys. Lett. B {\bf 115}, 482 (1982).
\bibitem{Campostrini86}
M. Campostrini, A. Di Giacomo, and S. Olejnik,
Z. Phys. C {\bf 31}, 577 (1986).
\bibitem{Simonov95}
Yu.A. Simonov, S. Titard, and F.J. Yndurain, Phys. Lett. B {\bf 354}, 435 
(1995).
\bibitem{Dosch87}
H.G. Dosch, Phys. Lett. B {\bf 190}, 177 (1987).
\bibitem{Dosch88}
H.G. Dosch and Yu.A. Simonov, Phys. Lett. B {\bf 205}, 339 (1988).
\bibitem{Simonov89}
Yu.A. Simonov, Nucl. Phys. B {\bf 324}, 67 (1989).
\bibitem{Nachtmann84}
O. Nachtmann and A. Reiter, Z. Phys. C {\bf 24}, 283 (1984).
\bibitem{Landshoff87}
P.V. Landshoff and O. Nachtmann, Z. Phys. C {\bf 35}, 405 (1987).
\bibitem{Kramer90}
A. Kr\"amer and H.G. Dosch, Phys. Lett. B {\bf 252}, 669 (1990).
\bibitem{Dosch94}
H.G. Dosch, E. Ferreira, and A. Kr{\"a}mer, Phys. Rev. D {\bf 50}, 1992 (1994).
\bibitem{Dorokhov97}
A.E. Dorokhov, S.V. Esaibegyan, and S.V. Mikhailov, Phys. Rev. D {\bf 56},
4062 (1997).
\bibitem{Ilgenfritz97}
E.--M. Ilgenfritz, B.V. Martemyanov, S.V. Molodtsov, M. M\"uller--Preussker,
and Yu.A. Simonov, Phys. Rev. D {\bf 58}, 114508 (1998).
\bibitem{Campostrini84}
M. Campostrini, A. Di Giacomo, and G. Mussardo, Z. Phys. C {\bf 25}, 173 (1984).
\bibitem{DiGiacomo92}
A. Di Giacomo and H. Panagopoulos, Phys. Lett. B {\bf 285}, 133 (1992).
\bibitem{npb97}
A. Di Giacomo, E. Meggiolaro, and H. Panagopoulos, Nucl. Phys. B {\bf 483}, 
371 (1997).
\bibitem{plb97}
M. D'Elia, A. Di Giacomo, and E. Meggiolaro, Phys. Lett. B {\bf 408}, 315 
(1997).
\bibitem{Brambilla98}
G. Bali, N. Brambilla, and A. Vairo, Phys. Lett. B {\bf 421}, 265 (1998).
\bibitem{Michael88}
C. Michael and M. Teper, Phys. Lett. B {\bf 206}, 299 (1988).
\bibitem{Jamin98}
M. Eidem\"uller and M. Jamin, Phys. Lett. B {\bf 416}, 415 (1998).
\bibitem{Nachtmann97}
O. Nachtmann, in {\it Perturbative and Nonperturbative Aspects of Quantum
Field Theory}, edited by H. Latal and W. Schweiger (Springer--Verlag,
Berlin, Heidelberg, 1997).
\bibitem{Nachtmann98}
E. Berger and O. Nachtmann, Heidelberg preprint, HD--THEP--98--33 (1998);
E--print Archive, hep--ph/9808320.
\bibitem{SVZ79}
M.A. Shifman, A.I. Vainshtein, and V.I. Zakharov, Nucl. Phys. B {\bf 147},
385; 448; 519 (1979).
\bibitem{Wilson69}
K.G. Wilson, Phys. Rev. {\bf 179}, 1499 (1969).
\bibitem{Mueller85}
A.H. Mueller, Nucl. Phys. B {\bf 250}, 327 (1985);\\
A.H. Mueller, in {\it QCD, 20 Years Later}, edited by P.M. Zerwas and
H.A. Kastrup (World Scientific, Singapore, 1993).
\bibitem{NSVZ81}
V.A. Novikov, M.A. Shifman, A.I. Vainshtein, and V.I. Zakharov, 
Nucl. Phys. B {\bf 191}, 301 (1981).
\end{thebibliography}
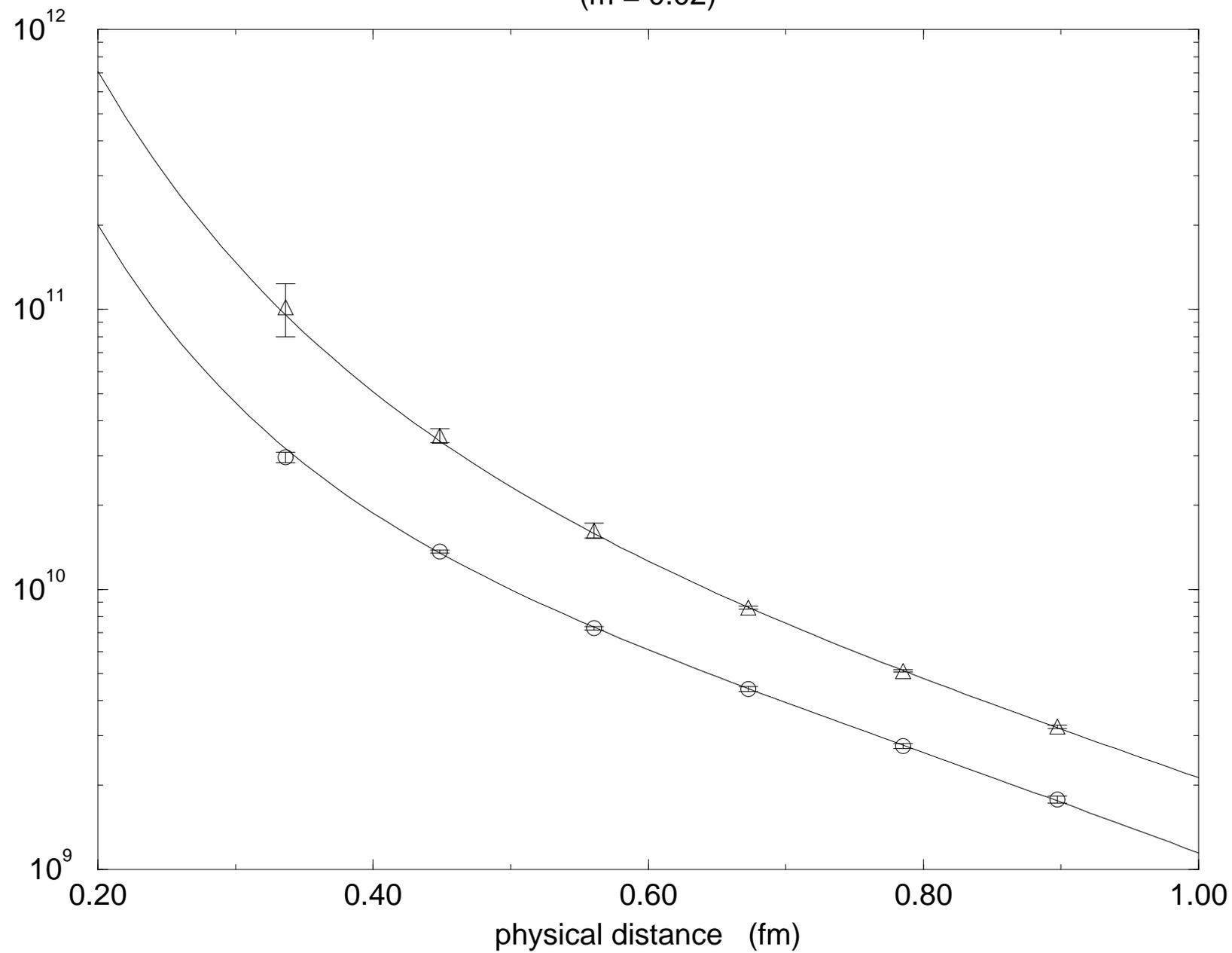
\end{document}